\documentclass[11pt]{article}

\usepackage{latexsym}

\def\thefootnote{\fnsymbol{footnote}}
\def\bea{\begin{eqnarray}}
\def\eea{\end{eqnarray}}
\def\beq{\begin{equation}}
\def\eeq{\end{equation}}

\def\rvev{\right\rangle}
\def\lvev{\left\langle}
\def\superint{\int d^{4}\theta}

\def\bM{\bar{M}}

\def\bC{\bar{C}}

\def\W{\overline{W}}

\def\G{{\cal G}}

\def\tG{{\tilde G}}

\def\hZ{{\widehat Z}}

\def\D{{\cal D}}

\def\bD{\bar{\D}}

\def\pp{\partial}

\def\ibar{\bar{\imath}}

\def\bbet{\bar{\beta}}
\def\[{\left [}
\def\]{\right ]}
\def\({\left (}
\def\){\right )}
\def\lbr{\left\{}
\def\rbr{\right\}}
\def\r{\right|}
\def\l{\left.}

\def\T{\bar{T}}

\def\z{\bar{z}}

\def\R{{\cal{R}}}

\def\STr{{\rm STr}}
\def\Tr{{\rm Tr}}

\def\Ti^{T^{(i)}}

\def\bR{\bar{R}}

\def\L{{\cal L}}
\def\R{{\cal{R}}}

\def\hZ{\hat{Z}}

\def\t{\bar{t}}

\def\hV{\hat{V}}
\def\n{\bar{n}}
\def\m{\bar{m}}

\def\bth{\bar{\theta}}

\def\A{\bar{A}}
\def\B{\bar{B}}

\def\Y{{\bar{Y}}}
\def\Z{{\bar{Z}}}

\def\bph{\bar{\phi}}
\def\bch{\bar{\chi}}
\def\bPh{\bar{\Phi}}

\def\bF{\bar{F}}

\def\bj{\bar{\jmath}}
\def\a0{\alpha_0}
\def\const{{1\over32\pi^2}}

\begin{document}

\begin{titlepage} 
\begin{center}
            \hfill    LBNL-45489 \\
            \hfill    UCB-PTH-00/12 \\
            \hfill hep-th/004170\\
	\hfill April 2000 
\vskip .1in {\large \bf Quantum-Induced Soft Supersymmetry Breaking In
Supergravity}\footnote{This work was supported in part by the
Director, Office of Science, Office of Basic Energy Services, of the
U.S. Department of Energy under Contract DE-AC03-76SF00098 and in part
by the National Science Foundation under grant PHY-95-14797.}\\[.1in]

Mary K. Gaillard and Brent Nelson

{\em Department of Physics, University of California, and 

 Theoretical Physics Group, 50A-5101, Lawrence Berkeley National Laboratory, 

      Berkeley, CA 94720, USA}

\begin{abstract}
	We calculate the one-loop quantum contributions to soft
supersymmetry breaking terms in the scalar potential in supergravity
theories regulated \`a la Pauli-Villars. We find ``universal''
contributions, independent of the regulator masses and tree level soft
supersymmetry breaking, that contribute gaugino masses and A-terms equal
to the ``anomaly mediated'' contributions found in analyses using
spurion techniques, as well as a scalar mass term not identified in
those analyses.  The universal terms are in general modified -- and in
some cases canceled -- by model-dependent terms.  Under certain
restrictions on the couplings we recover the one-loop results of
previous ``anomaly mediated'' supersymmetry breaking scenarios.  We
emphasize the model dependence of loop-induced soft terms in the
potential, which are much more sensitive to the details of Planck
scale physics then are the one-loop contributions to gaugino masses.
We discuss the relation of our results to previous analyses.

\end{abstract}
\end{center}
\newpage
\pagestyle{empty}
\null
\end{titlepage}
\newpage
\renewcommand{\thepage}{\arabic{page}}
\setcounter{page}{1}
\def\thefootnote{\arabic{footnote}}
\setcounter{footnote}{0}

There has been considerable interest recently in soft supersymmetry
breaking induced by quantum corrections, starting with the
observation~\cite{rs,hit} that there are several ``anomaly mediated''
contributions: a gaugino mass term proportional to the
$\beta$-function, an A-term proportional to
the chiral multiplet (matrix-valued) $\gamma$-function, and a scalar
mass term proportion to the derivative of the $\gamma$-function,
arising first at two-loop level.  This contribution to the gaugino
masses has been confirmed in subsequent calculations~\cite{gnw,bag}.
The result in~\cite{gnw} was found by an analysis of the relevant
loop-induced superfield operator in K\"ahler $U(1)$
superspace~\cite{bgg}, and also by an explicit Pauli-Villars (PV)
calculation.  The
``anomaly mediated'' A-term contribution has also been confirmed by a
Pauli-Villars calculation given in~\cite{pv2} as an illustrative
application of PV regularization of supergravity.  Here we extend
the PV calculation to obtain the one-loop contribution to scalar masses.
We also display our result in the form of a superfield operator, and 
indicate the origin of this operator as a superspace integral.
  We work with the
standard chiral formulation of supergravity in K\"ahler $U(1)$
superspace, with the Einstein term canonically normalized.  
The full contribution to gaugino masses in
string-derived supergravity models with the dilaton in a linear
supermultiplet, and including a Green-Schwarz term and string
threshold effects, was presented in~\cite{gnw}.  A general
parameterization of all the soft supersymmetry breaking terms in the
context of superstring-derived supergravity will be given
elsewhere~\cite{bgn}.

The points we wish to emphasize in the PV calculation given here are
1) the presence in general of $O(m_{\tG})$ contributions to the scalar
masses that are proportional to the chiral supermultiplet
gamma-function (rather than its derivative, which is a two-loop
effect), and 2) the difference between gaugino masses and soft terms
in the scalar potential with respect to dependence on the details of
Planck-scale physics.  To this end we will present our calculations
under the simplifying assumption that the Pauli-Villars squared-mass
matrix commutes with other operators that are relevant to quantum
corrections. The full PV mass-dependence in the general case will be
indicated in the final result. We further restrict our analysis to
one-loop order and retain only terms of lowest order in $m_{\tG}/m_P$,
where $m_{\tG}$ and $m_P$ are the gravitino mass and the reduced
Planck mass, respectively.  We then use our results to address the
issue of anomaly-mediated supersymmetry breaking~\cite{rs,hit,spurions,other}.

The one-loop logarithmic divergences of standard chiral
supergravity were determined in~\cite{vid}, and it was shown
in~\cite{pv2,mkg} that they can be regulated\footnote{The full
regulation of gravity loops requires the introduction of Abelian gauge
superfields as well; these play no role here and we ignore them.} by a
set of Pauli-Villars chiral superfields $\Phi^A$. As in these references
we denote the light superfields by $Z^i$, and introduce covariant 
derivatives of the superpotential $W(Z^i)$ as follows:
\bea A &=& e^KW, \quad A_i
= D_iA = \pp_iA, \quad A_{ij} = D_iD_jA = \pp_i\pp_jA -
\Gamma^k_{ij}A_k, \nonumber \\ \A^i &=& K^{i\m}\A_{\m}, \quad etc.,
\quad K_{i\m} = \pp_i\pp_{\m}K(Z^i,\Z^{\m}),
\quad \pp_i = {\pp\over\pp Z^i}, \label{nota}\eea
where $\Gamma^k_{ij}$ is the affine connection associated with the K\"ahler
metric $K_{i\m}$ and its inverse $K^{i\m}$.
In the regulated theory, the one-loop correction to the chiral multiplet 
K\"ahler potential is given by~\cite{pv2} 
the superfield operator (up to a Weyl transformation
necessary to put the Einstein term in canonical form\footnote{This
brings in terms with factors of $V_{tree}$ that we neglect since if
$<V_{tree}> = 0$, they can at most give small corrections to the tree
level soft terms.})  
\beq \Delta L = -
\const\superint E e^{-K} \sum_{AB}\eta_A\A^{AB}A_{AB}\ln(m_A^2/\mu^2) =
\superint E \delta K(Z^i,\Z^{\m}),\label{sfl}\eeq
where $A_{AB}(Z^i,\Z^{\m})$ is defined as in (\ref{nota}), 
with the light field 
indices $i,j$ replaced by PV indices $A,B$, $\eta_A = \pm 1$ is the
PV signature, $m_A$ is the (supersymmetric) PV mass, and $\mu$ is the 
(scheme-dependent) normalization point.  The wave function renormalization
matrix is given by 
\bea \gamma^j_i &=& \lvev K^{j\n}D_{\n}D_i{\pp\over\pp\ln\mu^2}\delta K\rvev =  
\const\lvev D^jD_i\sum_{AB}\eta_A(e^{-K}\A^{AB}A_{AB})\rvev \nonumber \\ &=&  
\const\lvev e^{-K}\sum_{AB}\eta_A\A^{jAB}A_{iAB}\rvev + \cdots,\eea
where here and throughout ellipses represent terms of higher dimension.

The regulation of matter and Yang-Mills loop contributions to the matter
wave function renormalization requires the
introduction of PV chiral superfields $\Phi^A = Z^I,Y_I,\varphi^a$, which
transform according to the chiral matter, anti-chiral
matter and adjoint representations of the gauge group and have signatures
$\eta_A = -1,+1,+1,$ respectively.  These fields 
couple to the light fields through the superpotential\footnote{Full regulation 
of the theory requires several copies of fields with the same gauge 
quantum numbers, and the coupling parameters and signatures
given here actually represent weighted average values.}
\beq W(\Phi^A,Z^i) = {1\over2}W_{ij}(Z^k)Z^IZ^J + \sqrt{2}\varphi^aY_I(T_aZ)^i 
+ \cdots, \label{pvpot}\eeq
where $T_a$ is a generator of the gauge group, and their K\"ahler potential 
takes the form
\beq K(\Phi^A,\bPh^{\A} ) = K_{i\m}Z^I\Z^{\bM} + K^{i\m}Y_I\Y_{\bM}
+ g^{-2}_ae^K|\varphi^a|^2 + \cdots,\label{pvkal} \eeq
where $g_a$ is the (possibly field-dependent) gauge coupling constant for the
gauge subgroup $\G_a$.
With these choices, the ultraviolet divergences
cancel, and, for the leading (lowest dimension) contribution,
one obtains the standard result for the matter wave function
renormalization in the supersymmetric gauge~\cite{barb}
\beq \gamma^j_i = \const e^K\sum_{AB}\eta_AW_{iAB}\W^{jAB} = \const\[
4\delta^j_i\sum_ag^2_a(T^2_a)^i_i - e^K\sum_{kl}W_{ikl}\W^{jkl}\] .\label{gam}\eeq  
The matrix (\ref{gam}) is diagonal
in the approximation in which generation mixing is neglected in the
Yukawa couplings; in practice only the $T^cQ_3H_u$ Yukawa coupling is
important.  We will make this approximation in the following, and
set 
\bea \gamma_i^j &\approx&\gamma_i\delta^j_i, \quad \gamma_i  = \gamma_i^W +
\gamma_i^g, \quad \gamma_i^W= \sum_{jk}\gamma_i^{jk},\quad
\gamma_i^g = \sum_a\gamma_i^a, \nonumber \\ \gamma_i^a &=& 4g^2_a(T^2_a)^i_i,
\quad \gamma_i^{jk} = -e^K(g_ig_jg_k)^{-1}\left|W_{ijk}\right|^2,\eea
where for gauge-charged fields $Z^i$ the K\"ahler metric is
\beq K_{i\bj} = g_i(Z^n)\delta_{ij} + O{|Z^i|^2}, \eeq
with the $Z^n$ gauge singlets.  

The Lagrangian (\ref{sfl}) generally contains soft supersymmetry
(SUSY) breaking terms, displayed below, that are proportional 
to those of the tree-level Lagrangian.  What are usually
referred to as ``anomaly mediated'' soft SUSY-breaking terms are 
finite contributions that are not remnants, like (\ref{sfl}), of
the ultraviolet divergences.  To evaluate such terms in the framework of PV
regularization, we must retain all contributions that do not vanish in the
limit $m^2_A\to\infty.$  Here we are interested in the scalar potential, 
given by
\bea\L &=& {i\over2}\int{d^4p\over(2\pi)^4}\STr\eta\ln\(p^2 - m^2 - H\)
\nonumber \\ &=& -\const\STr\eta\[\(hm^2 + {1\over2}g^2\)\ln(m^2) +
{1\over2}h^2\ln\({m^2\over\mu^2}\) + {H^3\over6m^2} - {H^4\over24m^4}\]
\nonumber \\ & & \qquad + O\({1\over m^2}\), \label{int}\eea 
where $H$ is the effective field-dependent squared mass with the supersymmetric
PV mass matrix $m^2$ separated out:
\beq H_{PV} = H + m^2, \quad H = h + g, \quad h \sim m^0, \quad g\sim
m^1. \label{hpv}\eeq
The terms in (\ref{int}) proportional to $\ln(m^2/\mu^2)$ are the bosonic
part of (\ref{sfl}).  The first term in (\ref{int}), 
proportional to $m^2$, is the remnant of the quadratically divergent
contribution~\cite{mkg}.  It is completely controlled by Plank scale physics, and can
be made to vanish with appropriate conditions on $m^2$.
If it is present, it contains A-terms and scalar masses proportional to 
the tree potential soft terms, with coefficients suppressed by 
$1/32\pi^2$; we 
neglect it in the following. 

The PV loops contribute soft SUSY breaking terms to the light field
effective Lagrangian if the PV tree Lagrangian contains such terms.
In the presence of SUSY breaking one generally expects the matrix $g$
in (\ref{hpv}) which is linear in $m$ to contain ``B-terms''.  Indeed
it is these B-terms that generate the ``anomaly mediated''
contributions to the gaugino masses and the A-terms of the light
theory\footnote{There may also be B-terms generated at one loop in the
light theory if there are quadratic holomorphic terms in its
tree-level superpotential or K\"ahler potential.  These contributions
were considered in~\cite{pv2}; we ignore them and use the expression
``B-term'' to designate the B-term proportional to the PV mass.} that
have been discussed in the literature~\cite{rs,hit,spurions,other}.  As we shall see
below, there are two contributions to SUSY-breaking scalar masses that
arise from a double B-term insertion in a Feynman diagram.  These two
contributions cancel, resulting in the assertion~\cite{rs,spurions,other} that
there is no anomaly mediated contribution to scalar masses at one
loop.  However, there can in general be soft masses and A-terms in the
matrix $h$ in (\ref{hpv}).  In leading order in $m_{\tG}^2/\mu^2$,
A-terms are present in the PV part of $h$ only if there are
dimension-three soft SUSY-breaking operators in the tree Lagrangian.
Soft PV mass terms, which in leading order contribute only to scalar
masses, are not similarly restricted by the low energy theory.
Specifically, if the regulator masses are constant there are always
soft squared-mass terms in the PV sector.

The PV mass for each
superfield $\Phi^A$ is generated by coupling it to a field
$\Phi^\alpha$ in the representation of the gauge group conjugate
to that of $\Phi^A$ through the superpotential term
\beq W_m = \sum_{(A,\alpha)}\mu_{A\alpha}\Phi^A\Phi^\alpha\label{pvmass} ,\eeq
where $\mu_{A\alpha}= \mu_{A\alpha}(Z^i)$ can in general be a
holomorphic function of the light superfields; we do not consider that
possibility here. If the K\"ahler potential for the PV fields is 
\beq K_{PV} = \sum_Xg_X(z)|\Phi^X|^2,\quad X = A,\alpha,\eeq
then the PV masses are
\beq m^2_A = m^2_\alpha = f_A\mu_{A\alpha}^2, \quad f_A = e^Kg^{-1}_Ag^{-1}_\alpha. 
\eeq
The general field-dependent matrix $H$ in (\ref{int}) has been evaluated
in~\cite{vid}.  Denoting by $H^{\chi}$ and $H^\phi$ the matrices $H$ in 
(\ref{hpv}) for fermions and bosons, respectively,
we have, with constant background fields,
\bea H^\chi &=& h^\chi + {r\over4}, \quad (h^\chi)^A_B
= e^{-K}A_{BC}\A^{AC}, \quad (h^\chi)^\alpha_\beta = 0,
\nonumber \\ (h^\chi)^\alpha_D &=& K^{\alpha\bbet}\mu_BK^{\B C}A_{CD} = 
g^\alpha_D = e^{-K}f_A\mu_{A\alpha}A_{AD} ,
\nonumber \\  (h^\chi)^A_\beta &=& \A^{AB}\mu_B = g^A_\beta, \quad
H^\phi = h^\chi + \hV + m_{\tG}^2 + R, \label{defh} \eea
where $m_{\tG}^2 = e^K|W|^2,\; \hV = e^{-K}A_j\A^{j} - 3m_{\tG}^2.$  
The last term in (\ref{defh}) depends
on the curvature of the PV metric:
\beq R^a_b = R^a_{b\m i}e^{-K}A^{\m}\A^i = - \delta_{ab}F^i\bF^{\m}
\pp_{\m}\(g^{-1}\pp_ig_a\) = - \delta_{ab}F^i\bF^{\m}
\pp_{\m}\pp_i\ln g_a, \quad a = A,\alpha,\label{pvr}\eeq
where $F^i = - e^{-K/2}\A^i$ is the auxiliary field of the supermultiplet $Z^i$.
Terms involving the space-time curvature $r$ are replaced by
terms proportional to the tree potential $V$ after a Weyl transformation
that restores the one-loop corrected Einstein term to canonical form. 
We assume throughout a vanishing cosmological constant,
$<V>=0$, so we can drop them.  Similarly, we can drop $\hV$
if D-terms vanish in the vacuum: $V = \hV + \D,\; \D = {1\over2}g^2
\sum_aD_a^2,\;(T^az)^iK_i, \; z^i = Z^i|, \; \langle D_a\rangle = 0$. 
Terms containing only powers of $h^\chi$ cancel in the supertrace,
so we get contributions only  from scalar trace terms that include
the scalar mass term $m_{\tG}^2 + R$ in (\ref{defh}) or 
factors of $H_{XY}= K_{X\bar X}H^{\bar X}_Y$:
\bea H_{AB} &=& h_{AB} = e^{-K}\(\A^iD_iA_{AB} - A_{AB}\A\), \quad
h_{\alpha\beta}= 0, \nonumber \\ 
H_{A\beta} &=& g_{A\beta} = e^{-K}\A^iD_i\(e^KW_{A\beta}\) - \A W_{A\beta}
= - \delta_{A\beta}\mu_{A\alpha}\(\A - \A^i\pp_i\ln f_A\).\eea
$H_{A\alpha}$ is the B-term mentioned 
above, and the part of $h_{AB}$ linear in $z^i - \langle z^i\rangle$ is
the A-term.  Neglecting B-terms in the tree Lagrangian, the leading 
contribution to $W_{AB}$ is linear in a gauge nonsinglet field $Z^i$.  
Explicitly expanding $H_{AB}$ gives 
\beq H_{AB} = e^KK^{i\bj}\W_{\bj}W_{iAB} - e^{K/2}F^n\pp_n\ln(g_ig_Ag_Be^{-K})
z^iW_{iAB}. \label{dac}\eeq
The second term in (\ref{dac}) is the A-term where $\langle F^n\rangle\ne0$ 
with $Z^n$ a gauge singlet in the SUSY-breaking sector.  
Assuming that the matter superpotienial is independent of $Z^n$,
the tree-level A-terms are given by
\bea  \lvev \pp_i\pp_j\pp_k V\rvev = 
a_{ijk}\lvev e^{K/2}W_{ijk}\rvev, \quad
a_{ijk} = \lvev F^n\pp_n\ln(g_ig_jg_ke^{-K})\rvev,\label{aijk}\eea
and the tree-level gaugino masses are given by
\beq m_a = {1\over2}\lvev F^n\pp_n\ln(g^{-2}_a)\rvev. \label{matree}\eeq
Using the couplings given in (\ref{pvkal}), we have
\bea F^n\pp_n\ln(g_ig_{Z^J}g_{Z^K}e^{-K}) = a_{ijk},
\quad  F^n\pp_n\ln(g_ig_{Y_I}g_{\varphi^a}e^{-K}) = 2m_a.\label{rels} \eea
Then we obtain
\bea \Tr\eta h^2 &\ni& 2\sum_{AB}\eta_Ah_{AB}h^{AB}\ni 
- 2e^{3K/2}W_jz^i\sum_{AB}\eta_AW_{iAB}
\W^{jAB}F^n\pp_n\ln(g_jg_Ag_Be^{-K}) + {\rm h.c.}
\nonumber \\ &=& -64\pi^2e^{K/2}W_iz^i\[\sum_{jk}\gamma_i^{jk}a_{ijk}
+ 2\sum_a\gamma_a^im_a\] + {\rm h.c.}, \label{h2a}\eea
for the leading contribution to the A-term from the 
second term in (\ref{int}), {\it i.e.}, the contribution from the
shift in the potential due to the shift in the K\"ahler potential.  
The leading order contribution to the ``anomaly-induced'' A-term arises
from a PV loop diagram with one B-term insertion:
\bea \Tr\eta{H^3\over6m^2} &\ni& \Tr\eta{ hg^2\over2m^2} \ni 
\sum_{AB}{\eta_A\over2m_A^2}h^A_{\B}\(g^{\B}_{\gamma}g^\gamma_A +
g^{\B}_{\bar{\gamma}}g^{\bar{\gamma}}_A\) + {\rm h.c.}\nonumber \\
&\ni& -32\pi^2e^{K/2}W_iz^i\[\gamma_im_{\tG} 
+ F^{n}\pp_{n}\(\sum^a\gamma^a_i\ln f_{ia}  + \sum_{jk}\gamma^{jk}_i\ln f_{jk}\)\]
 + {\rm h.c.}, \nonumber \\ f_{AB} &=& \sqrt{f_Af_B}, \quad 
f_{jk} = \sqrt{f_{Z^J}f_{Z^K}}, \quad f_{ia} = \sqrt{f_{\varphi^a}f_{Y_I}}, \eea
which reduces to the ``anomaly mediated term'' found in~\cite{hit} 
provided that $\langle F^{n}\pp_{n}\ln f_A\rangle = 0.$  We discuss 
below the circumstances under which this is the case.
The full leading-order A-term Lagrangian is
\bea \L_A &=&  e^{K/2}\sum_iW_iz^i\bigg[\gamma_im_{\tG}
+ \sum_a\gamma_i^a\(2m_a\ln(m^2_{ia}/\mu^2) + F^{n}\pp_{n}\ln f_{ia}\)\nonumber \\ & & 
+ \sum_{jk}\gamma_i^{jk}\(a_{ijk}\ln(m^2_{jk}/\mu^2) + F^{n}\pp_{n}\ln f_{jk}\)\bigg] 
+ {\rm h.c.} + \cdots,
\nonumber \\ m_{AB}^2 &=& m_{A}m_{B},\quad
m^2_{jk} = m_{Z^J}m_{Z^K}, \quad m^2_{ia} = m_{\varphi^a}m_{Y_I}. \label{aterm} \eea

Scalar masses get a contribution from the term quartic in $H$:
\bea \Tr\eta{H^4\over24m^2} &\ni& \Tr\eta{g^4\over24m^2} \ni \sum
{\eta_A\over12m_A^2}\[2\(g^A_{\beta}g^{\beta}_Cg^C_{\bar{\delta}}g^{\bar{\delta}}_A 
+ g^\alpha_Bg^B_\gamma g^\gamma_{\bD}g^{\bD}_\alpha\) +
g^A_{\beta}g^{\beta}_{\bC}g^{\bC}_{\bar{\delta}}g^{\bar{\delta}}_A
+ g^\alpha_Bg^B_{\bar{\gamma}}g^{\bar{\gamma}}_{\bD}g^{\bD}_\alpha\]
\nonumber \\ & & + {\rm h.c.}\ni e^{-K}\sum_{AB}\eta_A|m_{\tG} + F^i\pp_i\ln
f_A|^2A_{AB}\A^{AB},\label{h4} \eea 
which corresponds to two B-term insertions in the PV loop.  
The contribution from the cubic term is
\bea \Tr\eta{hg^2\over2m^2} &\ni&
{1\over2}\sum_{AB}\eta_Ah^A_{\B}\[g^{\B}_\beta g^\beta_A{1\over m^2_B} +
g^{\B}_{\bar{\alpha}} g^{\bar{\alpha}}_A{1\over m^2_A}\] \nonumber \\
& & + {1\over2}m^2_{\tG}\sum_{AB}\eta_Ag^A_\beta g^\beta_A
\({1\over m^2_B} + {1\over m^2_A}\) \nonumber \\ & & -
{1\over2}\sum_{AB}\eta_Ag^A_\beta g^\beta_A \({1\over
m^2_B}F^n\bF^{\m}\pp_n\pp_{\m}\ln g_\beta + {1\over
m^2_A}F^n\bF^{\m}\pp_n\pp_{\m}\ln g_A\) \nonumber \\ & & +
{1\over2}\sum_B\eta_Bm_B^{-2}g_{B\beta}g^{B\beta}h^B_B + {\rm h.c.}
\nonumber \\ &\ni& {1\over2}e^{-K}\sum_{AB}\eta_A\[
F^n\pp_n\ln\(g_ig_Ag_Be^{-K}\)\]\(m_{\tG} + \bF^{\m}\pp_{\m}\ln
f_B \)z^iA_{iAB}\A^{AB} \nonumber \\ & & + {1\over2}
e^{-K}\sum_{AB}\eta_AA_{AB}\A^{AB}\(F^n\bF^{\m}\pp_n\pp_{\m}\ln f_A 
- m_{\tG}^2\) \nonumber \\ & & +
{1\over2}e^{-K}\sum_{AB}\eta_AA_{AB}\A^{AB}|m_{\tG} + \bF^{\m}\pp_{\m}\ln
f_B |^2 + {\rm h.c.},\label{h3}\eea 
where we used the vacuum condition
\beq \hV = K_{n\m}F^n\bF^{\m} - 3m^2_{\tG} = 0.\label{vac}\eeq
The last term in (\ref{h3}) is a double B-term insertion; it cancels 
(\ref{h4}) 
in the Lagrangian (\ref{int}).  The first term on the right hand side of
(\ref{h3}) corresponds to one B-term and one A-term insertion, and the second
term corresponds to a PV soft squared-mass insertion.  Explicitly,
\beq F^n\bF^{\m}\pp_n\pp_{\m}\ln f_A - m_{\tG}^2 = \mu^2_A + \mu^2_\alpha,
\label{rels2}\eeq
where
\beq \mu^2_a = m^2_{\tG} - F^n\bF^{\m}\pp_n\pp_{\m}\ln g_a, \quad a = A,\alpha \label{mtree}\eeq
is the soft SUSY-breaking squared mass of the field $\Phi^a$.  For $a=A$
the masses are determined by the SUSY-breaking masses of the tree Lagrangian
\beq \mu^2_{Z^I} = \mu^2_{z^i} \equiv \mu^2_i, \quad \mu^2_{Y_I} = 2m_{\tG}^2
- \mu^2_i, \quad  \mu^2_{\varphi^a} = - 2m^2_{\tG} - m^2_a, \label{rels3}\eeq
so these terms give no contribution if $\mu_i=m_a=0$.  However, even if no
soft SUSY-breaking masses are present in the tree Lagrangian, one cannot 
{\it a priori} exclude such terms in the theory above the effective cut-off, that
could be reflected in soft SUSY breaking masses $\mu^2_\alpha$ 
in the PV sector that parameterizes
the underlying Planck scale physics.  Finally we have 
\bea \Tr h^2 &\ni& 2\sum_{AB}\eta_A\(H^A_BH^B_A + H_{AB}H^{AB}\) 
\nonumber \\ &\ni&
2e^{-K}\sum_{AB}\eta_A\bigg[2\mu^2_A\W^{AB}W_{AB} \nonumber \\ 
 & & + z^i\z^{\bj}F^n\bF^{\m}\(\pp_n\ln(g_ig_Ag_Be^{-K})\)\(\pp_{\m}
\ln(g_jg_Ag_Be^{-K})\)\W_{\bj}^{AB}W_{iAB}\bigg],\label{h2}\eea
for the part of the renormalization of the K\"ahler potential
that contributes to scalar masses.  Using (\ref{rels}) and 
(\ref{rels2})--(\ref{rels3}), the full scalar mass term is
\bea \L_m &=& \sum_i|z^i|^2\Bigg(m^2_{\tG}\gamma_i 
\nonumber \\ & &
+ \sum_a\gamma^a_i\lbr\(3m_a^2 - \mu^2_i\)\ln(m^2_{ia}/\mu^2) + 
F^n\bF^{\m}\pp_{\m}\pp_n\ln f_{ia} + m_a\[\(\bF^{\m}\pp_{\m} + F^n\pp_n\)
\ln f_{ia} + 2m_{\tG}\]\rbr\nonumber \\ & &
+ \sum_{jk}\gamma^{jk}_i\bigg\{\(\mu_j^2 + \mu^2_k + a^2_{ijk}\)
\ln(m^2_{jk}/\mu^2) + 
F^n\bF^{\m}\pp_{\m}\pp_n\ln f_{jk}\nonumber \\ & & \qquad\qquad
+ {1\over2}a_{ijk}\[\(\bF^{\m}\pp_{\m} + F^n\pp_n \)\ln f_{jk} + 2m_{\tG}\]
\bigg\}\Bigg) + \cdots. \label{mass}\eea
In the absence of tree-level soft SUSY breaking, this expression reduces
to the first (``universal'') term if $\lvev F^n\bF^{\m}\pp_n\pp_{\m}f_A\rvev=0
$. 

The above results hold in the general case of a noncommuting 
squared-mass matrix, with the replacements
\bea \ln m^2_{AB} &=& q(m^2_A,m^2_B) = {m^2_A\ln(m^2_A/\mu^2) -
m^2_B\ln(m^2_B/\mu^2)\over m^2_A - m^2_B} - 1,\nonumber \\ 
\pp_{\m}\ln f_{AB} &=& \pp_{\m}q(m^2_A,m^2_B),\nonumber \\ 
\pp_n\pp_{\m}\ln f_{AB} &=& \pp_n\pp_{\m}q(m^2_A,m^2_B).\label{q}\eea
The soft SUSY-breaking Lagrangian for the canonically normalized scalars
$\phi_R$ is obtained by making the substitution $\phi^i=g_i^{-{1\over2}}
\phi^i_R$ in $\L_A + \L_m$.  

The Lagrangian $\L_A + \L_m$ is the bosonic part of the superfield
Lagrangian
\beq \L_1 = \const\superint E\sum_{AB}\eta_Ae^{K/2}\W^{AB}\[
\ln\Box_\chi - \bR{6\over\Box_\chi}R 
- q(m_A^2,m^2_B)\]e^{K/2}W_{AB} + \cdots, \label{loop}\eeq
where~\cite{bgg} $R$ is a superfield constructed from elements of the super-Riemann 
tensor, $\lvev 2R|\rvev = m_{\tG}$, and
\beq \Box_\chi = {1\over16}\(\bD^2 - 8R\)\(\D^2 - 8\R\) \eeq
is the chiral superfield propagator given in~\cite{1001}. Provided the K\"aher
metric is defined from the full K\"ahler potential including the PV part,
$K_{AB}$ is covariantly constant, and $\phi = e^{K/2}W_{AB}$ is a chiral
superfield~\cite{bgg} of chiral weight $+2$:
\beq \lvev\Box_\chi\phi\rvev = \lvev\(\Box + {1\over2}R\D^2 + 4R\bR\)\Phi\rvev
,\label{box}\eeq
where we used the vacuum condition (\ref{vac}).
The superfield $f(\Box_\chi)\phi$ is also a chiral
superfield of chiral weight 2, since $(\bD^2 - 8R)$
is the weight-zero chiral projection operator.  Evaluating (\ref{loop})
with the methods of~\cite{bgg}:
\beq\superint \Phi = {1\over16}\l\(\D^2 - 24\bR\)\(\bD^2 - 8R\)\Phi\r + 
{\rm gravitino \;terms},\eeq
and expanding in inverse powers of the d'Alembertian, 
$\langle\Box\rangle = \mu^2$,
we recover the scalar potential given in (\ref{aterm}), (\ref{mass}) and
(\ref{q}), up to corrections of order $m^2_{\tG}/\mu^2$.
To understand the origin of the expression (\ref{loop}),
consider the tree-level superfield Lagrangian for quantum fluctuations
$\hZ$ around canonically normalized background superfields $Z$:
\beq \L_0 = \superint E\[\sum_i|\hZ^i|^2 + e^{K/2}\({1\over4R}\hZ^i\hZ^jW_{ij}(Z) 
+ {\rm h.c.}\) \]. \label{tree}\eeq
Variation of the action $S = \int d^4x\L_0$ with respect to the unconstrained 
superfields $\rho^I$, defined by
\bea \rho^I = \pmatrix{\rho^i\cr\rho^{\ibar}\cr}, \quad \hZ^i = {1\over4}
\(\bD^2 - 8R\)\rho, \quad \rho^{\ibar} = \(\rho^i\)^{\dag}, \eea
gives the inverse superfield propagator
\bea \Delta^{-1}_{IJ}(y,y') &=& {\pp^2S\over\pp\rho^I(y)\pp\rho^J(y')}, \nonumber \\
\Delta^{-1}(y,y') &=& \pmatrix{\Box_{\bch}& -{1\over4}\bph\(\D^2-8\bR\)\cr
-{1\over4}\bph\(\bD^2-8R\)&\Box_\chi\cr}_y\delta^8(y-y'),\label{prop}\eea
where $y= x^m,\theta_\mu,\bth_{\dot\mu},$ and $\phi = e^{K/2}W_{ij}$ is a weight-2
chiral superfield.  In the flat superspace limit with $\phi = W_{ij} = {1\over2}m_{ij}$,
(\ref{prop}) reduces to the inverse of the Wess-Bagger free superfield propagator~\cite{wb}.
The effective one-loop action is determined by evaluating $\STr\ln\Delta$ in superspace.
Writing
\bea \Delta^{-1} &=& \Delta_0^{-1}\(1 + \delta\), \quad \Delta_0^{-1}(y,y') = 
\pmatrix{\Box_{\bch}& 0\cr 0&\Box_\chi\cr}_y\delta^8(y-y'),\nonumber \\  \delta(y,y') &=&
-{1\over4}\pmatrix{0& \Box_{\bch}^{-1}\bph\(\D^2-8\bR\)\cr
\Box_\chi^{-1}\bph\(\bD^2-8R\)&0\cr}_y\delta^8(y-y'),
\nonumber \\ \STr\ln\Delta &=& \STr\[\ln\Delta_0 + \ln(1 + \delta)\],
\label{exp}\eea
we are interested in the term proportional to $\delta^2$ in the expansion
of the log. Consider first the flat superspace limit:
$R=0,\;\Box_\chi\to\Box.$  Since $\delta(\theta) = \theta$,
one factor of $\delta^4(\theta - \theta')$ is removed by integration over 
$d^4\theta'$, and the other is removed by the spinorial derivative 
operators $\D^2,\bD^2$. The $x'$ integration can be performed by replacing
$\delta^4(x-x')$ by its Fourier
transform, yielding the flat space limit of the first term\footnote{This is 
obvious for the contribution proportional to $\gamma^W_i$, since 
$W_{Z^IZ^J} = W_{ij}$. The contribution proportional to $\gamma^g_i$ does
not arise from superpotential couplings, but it must be of the same form since
the result obtained from loops of massless fields is the same as that from the PV fields
$\varphi^a,Y_I$ with the superpotential couplings (\ref{pvpot}).} in (\ref{loop}).
In the curved superspace generalization of that term, the superdeterminant $E$
of the supervielbein appears as the Jacobian relating tangent space to 
Einstein superspace coordinates.  Additional spinorial derivatives appear in 
the expansion of $\Box_\chi^{-1}$ in powers of $\Box^{-1}$ [{\it c.f.} 
(\ref{box})].  For example, there is a contribution in which two $\theta$ 
factors in $\delta^2$ are removed by $\D^2$ in the numerator, and two others 
by a $\bD^2$ in the expansion of the denominator, resulting in a term
proportional to the second term in $(\ref{loop})$.  Other contributions
to this term arise from the superspace curvature implicit in the definition~\cite{bgg}
of the tangent space spinorial covariant derivative $\D_\alpha$.
The last term in (\ref{loop}) is obtained by replacing $\Box_\chi\to\Box_\chi
- m^2$ in the expression (\ref{exp}) for $\delta$ and dropping terms of order $m^{-2}$.

For completeness and comparison we give the result for the one-loop induced
(left-handed) gaugino mass~\cite{gnw,bag} under the same assumptions used here
to calculate $\L_A+\L_m$:
\bea \Delta m_a &=&  -{g_a^2(\mu)\over16\pi^2}\[(3C_a - C_a^M)m_{\tG} +
\sum_X\eta_XC_a^XF^n\pp_n\ln f_X\] +\cdots \nonumber \\ &=& 
-{g_a^2(\mu)\over16\pi^2}\[(3C_a - C_a^M)m_{\tG} 
+ C_a F^n\pp_nK - \sum_iF^nC_a^i\pp_n\ln f_i\] +\cdots, \nonumber \\
f_i &=& g_i^{-2}e^K,\label{delm}\eea
where $C_a,C_a^X,C_a^i$ are the quadratic Casimirs in the adjoint of the
gauge subgroup $\G_a$ and in the representations of $\Phi^X,Z^i$, respectively,
with $C_a^M = \sum_iC^i_a.$  The second equality in (\ref{delm}) follows
from the requirements of finiteness~\cite{mkg} and supersymmetry~\cite{tom}
of the chiral/conformal anomaly proportional to the squared gauge field
strength.  In contrast to the results in (\ref{aterm}) and (\ref{mass}), the 
leading one-loop contribution to the gaugino masses is completely determined
by the low energy theory.  In this case all gauge-charged PV fields $\Phi^X$ 
contribute, their mass matrix is block diagonal and commutes with the relevant 
operators, and the gauge-charge weighted masses are constrained
to give the second equality in (\ref{delm}).
On the other hand, only a subset of charged PV
fields contribute to the renormalization of the K\"ahler potential.
While the K\"ahler metrics of the fields $\Phi^A$ that appear in 
$W(\Phi^A,Z^i)$, Eq. (\ref{pvpot}), are determined as in (\ref{pvkal})
by the finiteness requirement, the metrics of the fields
$\Phi^\alpha$ to which they couple in $W_m$, Eq. (\ref{pvmass}),
are arbitrary.  Since the conformal anomaly associated with the renormalization
of the K\"ahler potential is a D-term, it is supersymmetric by
itself and there is no constraint analogous to the conformal/chiral
anomaly matching in the case of gauge field renormalization with
an F-term anomaly.  As a consequence the ``nonuniversal''
terms appearing in $\L_A + \L_m$ cannot be determined precisely in the
absence of a detailed theory of Planck scale physics.

As a check of our results, consider the ``no-scale'' model defined by
\beq K = k(S) + G, \quad G = -3\ln(T + \T - \sum_i |\Phi^i|^2), \quad
W = W(\Phi^i) + W(S), \label{ns}\eeq which has no soft SUSY breaking
in the observable sector $\Phi^i$ at tree level.  If we regulate the
theory so as to preserve the no-scale structure, we have $g_\alpha =
g_A$, and \beq f_A = f_i = h_i(s)e^{G/3}. \eeq  
Then 
\beq \pp_t\ln f_A = {1\over3}G_t,
\quad \pp_{\t}\pp_t\ln f_A = {1\over3}G_{t\t}.\eeq 
Vanishing vacuum energy at tree level requires $F^s=0$, so if 
$<\phi^i>=0,\;\phi^i=\Phi^i|$, the no-scale K\"ahler potential satisfies 
\beq - F^nG_n =
e^{K/2}WG_{\t}K^{\t t}G_t = 3m_{\tG}, \quad F^n\bF^{\m}G_{n\m} =
e^{K/2}|W|^2G_{\t}K^{\t t}G_t = 3m_{\tG}^2,\eeq 
and all the soft
SUSY-breaking terms, (\ref{aterm}), (\ref{mass}) and (\ref{delm}),
cancel,\footnote{The vanishing of the one-loop contribution to the
gaugino mass in this model was noted by Randall and Sundrum, private
communication.} in agreement with explicit calculations~\cite{sally}
and nonrenormalization theorems~\cite{np} in the context of this
model.  In the context of string theory however, the ``no-scale''
regularization is unacceptable, because it leads to a loop-corrected
Lagrangian that is anomalous under modular transformations, which are
known to be perturbatively unbroken in string theory.  Therefore one
must restore modular invariance by including, for example, a modular
covariant field dependence in the PV mass parameters in
(\ref{pvmass}), $\mu_{A\alpha} = \mu_{A\alpha}(T)$, which might
reflect string loop threshold corrections; since
these necessarily break the no-scale structure of the theory soft SUSY
parameters would be generated since $F^T\ne0$ in this toy model.
Alternatively, as shown in~\cite{pv2}, this theory can be regulated
with field-independent masses for the PV fields that contribute to the
renormalization of the K\"ahler potential: $\pp_n f_A =$ constant, in
which case the A-terms are precisely those found
previously~\cite{rs,hit,spurions,other}, and scalar masses are also generated
at one loop: $\Delta \mu^2_i = \gamma_im^2_{\tG}$.  Gauginos remain
massless, since their masses are insensitive to the specific choice of PV
regulator masses.

Randall and Sundrum~\cite{rs} considered a class of models defined by a 
K\"ahler potential
\beq K = - 3\ln\[1 - \sum_i|\Phi^i|^2 - f(Z^n,\Z^{\m})\],\label{rsk}\eeq
where the $\Phi^i$ represent gauge-charged matter, and the $Z^n$ are in a 
hidden
sector where SUSY is broken: $\langle\Phi^i\rangle = \langle F^i\rangle = 0,$ 
$\langle F^n\rangle \ne 0$.  For these models 
\beq \lvev K_{i\bj}\rvev = \lvev e^{K/3}\delta_{ij}\rvev = g_i\delta_{ij},
\quad \mu^2_i = a_{ijk} = 0,\eeq
from the definitions (\ref{aijk}) and (\ref{mtree}) and the vacuum condition
(\ref{vac}).  In addition there is no dilaton: $g_a =$ constant, $m_a = 0$, so
there is no soft SUSY breaking in the tree Lagrangian.  If we {\it assume}
$g_\alpha = g_A$, there are also no soft SUSY breaking parameters in the 
PV Lagrangian.  Then the scalar masses vanish at one loop, and we obtain:
\bea \ln f_{jk} &=& g_a^{-1}\ln f_{ia} = \ln f_i = K/3, \nonumber \\
\L_A &=& -\const e^{K/2}W_iz^i\gamma_i\(m_{\tG}
+ {1\over3}F^nK_n\) + {\rm h.c.} + \cdots, \nonumber \\ \Delta m_a &=&
-{g_a^2(\mu)\over16\pi^2}(3C_a - C_a^M)\(m_{\tG} + {1\over3}F^nK_n\) + \cdots.
\label{rs}\eea
To determine the model-dependent contribution proportional to $\lvev F^nK_n
\rvev$, we study the vacuum conditions $\lvev V\rvev = \lvev V_z\rvev =0$
for the potential $V(z=Z|)$ derived from $W(Z)$ and $K(Z) = - 3\ln[1- f(Z)]$, 
with the gauge-charged fields $\Phi$ set to zero.  This potential is 
classically invariant under the K\"ahler transformation 
\beq K(Z)\to K^\xi(Z) = K(Z) + \xi\ln|W(Z)|^2,\quad f^\xi = 
(1- f)|W|^{2\xi\over3},\quad W(Z)\to W^{1-\xi}(Z).  \label{kahlt}\eeq
If one imposes a ``separability'' condition~\cite{rs} on the superpotential
\beq W = W(\Phi^i) + W(Z^n),\label{rspot}\eeq
the redefinition (\ref{kahlt}) is not a classical invariance of the full theory
with $\Phi\ne 0$,  but rather defines a one-parameter family of models of the
general form defined by (\ref{rsk}) and (\ref{rspot}), with the same vacuum,
but with different couplings of the hidden sector to gauge-charged matter.

If $f(Z,\Z) = f(|Z|^2)$ and $\lvev z\rvev=\lvev Z|\rvev =0$, then 
$\lvev K_z\rvev
= \lvev3f'(z)\z\rvev = 0$, and the ``anomaly mediated'' results are recovered
in the dimension-three soft operators (\ref{rs}).  An example of this type
is given in~\cite{rs} for the case of a single hidden sector field $Z$.  
For the family of models generated from that one by the redefinitions
(\ref{kahlt}), we obtain for the coefficient of the soft terms in (\ref{rs}):
\beq B^\xi = m_{\tG} + {1\over3}\lvev F^zK^\xi_z\rvev = m_{\tG} + {\xi\over3}
\lvev F^z{W_z\over W}\rvev = (1-\xi)m_{\tG},\eeq
since $\lvev F^z\bF^{\z}K_{z\z}\rvev = -\lvev F^z(K_zW + W_z)\rvev = 
3m^2_{\tG}$ is invariant under (\ref{kahlt}).  As a second example,
consider the simpler K\"ahler potential, $f(Z) = |Z|^2$, with $W(Z) =
\lambda(1 +Z)^3$, which for $\Phi=0$ is classically equivalent, by a field
redefinition and a K\"ahler transformation, to the
no-scale theory defined by (\ref{ns}) with 
$f(T) = 1 - T - \T, \; T = (1-Z)/[2(1+Z)],\; W_T = 0$. In this case we find
\beq B^\xi = m^2_{\tG}\(1 - h^\xi\),\quad \xi \le h^\xi =
\lvev{\z + \xi + |z|^2(1-\xi)\over1+z}\rvev \le1,\eeq
if $0\le\xi\le1$, since $\lvev z\rvev$ is undetermined in this no-scale model,
but satisfies $|z|\le1$. For $\lvev z\rvev =\xi=0$, we have $\lvev F^zK_z
\rvev = 0$, giving the standard result~\cite{rs,hit,spurions,other} 
$B^\xi=m_{\tG}.$
For $\xi=1$, this model is precisely the one defined by (\ref{ns}), with 
$B^\xi=0$.  Quite generally, if $\lvev F^nW_n\rvev = 0$ in the 
class of models defined by the separability conditions (\ref{rsk}), 
(\ref{rspot}) and $g_\alpha = g_A$ for the PV fields, the soft SUSY-breaking
terms all vanish, since in this case $\lvev F^nK_n\rvev = \lvev F^n\bF^{m}
K_{i\m}\rvev/m_{\tG} = - 3m_{\tG}$ by the vacuum condition (\ref{vac}).

To summarize, we have found that the ``anomaly mediated'' results for
soft SUSY breaking rest on the separability assumptions stated above,
but also on more specific assumptions on the form of the hidden sector
potential.  We now address the question as to why these same results
were obtained by spurion analyses.\footnote{The authors
of~\cite{other} also pointed out that these results are correct only if
$\lvev F^nK_n\rvev$ can be neglected.} In its original 
incarnation~\cite{spurions},
these techniques of deriving observable sector soft SUSY breaking
terms were applied solely to models in flat superspace (such as models 
of gauge-mediated SUSY breaking). In these cases the K\"ahler
potential and superpotential obeyed the separability conditions
between observable and hidden sectors
\begin{eqnarray}
K_{\rm tot}=K_{\rm obs}\(\Phi,\bPh\)+K_{\rm hid}\(Z,\Z\), \quad
W_{\rm tot}=W_{\rm obs}\(\Phi\)+W_{\rm hid}\(Z\).
\label{flatsep}
\end{eqnarray}
Furthermore,
the observable sector K\"ahler potential was of a minimal variety: $K_{\rm
  obs}\(\Phi,\Phi\)=\sum_{i}|\Phi^{i}|^2$.

The key properties of models in which the leading contributions to
soft terms arise from the conformal anomaly were enumerated by Randall 
and Sundrum and are encapsulated in the form of the K\"ahler
potential~(\ref{rsk}). This K\"ahler potential was the result of
demanding separability in the function $\Omega = -3e^{-K/3}$:
\begin{eqnarray}
\Omega_{\rm tot}=-3+\Omega_{\rm obs}\(\Phi,\bPh\)+\Omega_{\rm hid}\(Z,\Z\),
\quad W_{\rm tot}=W_{\rm obs}\(\Phi\)+W_{\rm hid}\(Z\),
\label{rssep}
\end{eqnarray}
where the factors $-3$ ensure the canonical normalization of the
Einstein term in the supergravity Lagrangian. The separability
condition~(\ref{rssep}) and the requirement that $\Omega_{\rm obs}
\propto\sum_{i}|\Phi^{i}|^2$ (necessary to ensure vanishing tree level soft
SUSY breaking in the visible sector) give rise to~(\ref{rsk}). Of 
course in the flat space limit K\"ahler separability~(\ref{flatsep})
and separability in $\Omega$~(\ref{rssep}) are equivalent statements. Thus
the K\"ahler potential assumed 
in~(\ref{rsk}) is of precisely the limited class of potentials for
which the flat-space spurion techniques
can be imported into a supergravity context, as in Refs.~\cite{rs}
and~\cite{other}, without complication. This intimate connection
between~(\ref{rsk}) and the canonical flat space
of the spurion technique is not surprising as the ansatz
of~(\ref{rssep}) represents a set of models with very special
conformal properties, as we will elucidate below.

For dimension-three soft terms 
the distinction between curved and flat superspace is irrelevant, and
the dependence of the
anomaly-induced soft terms~(\ref{rs}) on the auxiliary multiplet of
supergravity is fixed by the conformal properties of the operators
involved~\cite{hit}. The complete anomaly contribution for the
dimension-two soft terms
given in~(\ref{mass}) not proportional to the normal
logarithmic running can in fact be obtained from the spurion
technique by use of the following construction. We promote the wave
function renormalization coefficient $Z$ to a spurion superfield
$\cal{Z}$ as in~\cite{spurions}. However, this field is not only dependent on the
chiral compensator $\eta = 1+F_{\eta}\theta^{2}$ and its Hermitian
conjugate, but also on a {\em real} superfield. Using the PV soft term 
definitions in~(\ref{rels}) we can
see that this spurion is given schematically by 
\begin{equation}
V=1-(a+2m_{a})\theta^2 - (a+2m_{a}){\bth}^2 + \mu_{a}^{2}\theta^2{\bth}^{2},
\label{PVspur}
\end{equation}
where here $a$ and $2m_{a}$ generically represent the tree-level
A-terms of the Pauli-Villars sector that correspond, respectively, to the
tree-level A-terms $a$ and gaugino masses $m_a$ of the light field sector,
and $\mu_{a}$ represents the soft scalar masses of the Pauli-Villars
sector. The functional dependence of the superfield
${\cal Z}$ on these spurions is given by 
\begin{equation}
{\cal Z}={\cal Z}\(\frac{\mu {\rm V}}{({\eta}{\bar{\eta}})^{1/2}}\)
\end{equation}
with $\mu$ the renormalization point as before, in analogy with
Ref.~\cite{spurions}.

We can now perform a Taylor series expansion of this expression about
$\theta=0$ to obtain
\begin{eqnarray}
\ln{\cal Z}&=&\ln{Z(\mu)}+
\frac{\partial\ln{Z(\mu)}}{\partial\ln{\mu}}\[-\(a+2m_{a}+\frac{F_{\eta}}{2}\)\theta^{2} 
- \(a+2m_{a}+\frac{\bar{F}_{\eta}}{2}\)\bth^{2}\] \nonumber \\
 & &+ \frac{\partial\ln{Z(\mu)}}{\partial\ln{\mu}}\[\(\mu_{a}^{2}
+\frac{(a+2m_{a})F_{\eta}}{2} +\frac{(a+2m_{a})\bar{F}_{\eta}}{2}
+\frac{|F_{\eta}|^2}{4}\)\theta^{2}\bth^{2}\] + {\rm two-loop}.
\label{rsexpand}
\end{eqnarray}
Now the {\em chiral} field redefinition
\begin{equation} 
\eta \to \eta'=
Z(\mu)^{1/2}\exp{\(-\frac{1}{2}\frac{\partial\ln{Z(\mu)}}{\partial\ln{\mu}}F_{\eta}\theta^{2}\)}\eta
\label{rotate}
\end{equation} 
can be
performed as usual in the spurion derivation to eliminate the one-loop
contribution to soft
masses arising from the supergravity auxiliary field and generate the
one-loop contribution to the A-terms. This process is equivalent 
to the cancellation of the double B-term insertions mentioned
above~(\ref{rels2}). Note 
the importance of the assumptions of~(\ref{rsk}), in particular the
fact that the K\"ahler potential for the observable sector is minimal
to lowest order in $1/m_{\rm P}$, for the rotation~(\ref{rotate}) to be 
performed. 

This same chiral rotation cannot be performed on the real
superfield contributions of the Pauli-Villars soft SUSY breaking
terms. This real superfield is not itself the product of a chiral and
anti-chiral superfield. The terms proportional to $\theta^2$ are thus irrelevant
provided that there is no SUSY breaking in the observable sector. This 
is a result of the celebrated holomorphy that underlies the spurion technique. The
scalar masses are then read off from the $\theta^{2}\bth^{2}$
component of~(\ref{rsexpand}). Use
of~(\ref{rels2}) and the equation of motion for the auxiliary field
$F_{\eta}$ then leads to identification with~(\ref{mass}).

The question remains, why do flat-space spurion techniques imply the
vanishing of the Pauli-Villars tree-level soft SUSY breaking parameters
independent of the specific nature of the K\"ahler potential and
superpotential? The answer can again be found in the special class of
supergravity theories for which these techniques can be
applied. Specifically, as mentioned above a ``sequestered'' sector
model is really nothing more than a model on an Einstein-K\"ahler
manifold, of which the no-scale models are a particular
subset~\cite{noscale}. These
spaces are defined by the fact that the curvature is proportional to
the metric. The constant of proportionality determines the
normalization of the Einstein term in the supergravity lagrangian. In
flat space these are empty statements, but in curved space a properly
normalized Einstein term in an Einstein-Kahler manifold will cause the 
scalar mass term $R+m_{\tG}^{2}$ in~(\ref{defh}) to be proportional to
$\hV$. Hence in a space with vanishing cosmological constant and
K\"ahler potential given by~(\ref{rsk}) (which is equivalent to using
the spurion technique in flat space) the PV tree-level soft masses are
identically zero. It follows that no one-loop scalar masses will be
generated in these theories by the conformal anomaly.

In conclusion, we have shown that, even in the absence of soft SUSY
breaking at tree level, the loop-induced soft SUSY breaking operators
are not uniquely determined by anomalies.  In particular, the hidden
sector K\"ahler potential and superpotential must be separately
specified.  This is closely related to the well-known
fact\footnote{For a recent discussion of this and related issues,
see~\cite{bag2}.} that K\"ahler invariance of supergravity is broken
at the quantum level, as is manifest in the expressions (\ref{rs});
$F^n$ is K\"ahler covariant while $K_n$ is not.  Once the full low
energy Lagrangian is specified, including any hidden sector, the
one-loop gaugino masses are completely determined by the requirements
of finiteness and supersymmetry of the K\"ahler anomaly.  However the
soft terms in the {\mbox scalar} potential depend on the details of Planck
scale physics, since the corresponding PV couplings are not
sufficiently constrained.  In particular, scalars can acquire masses
at one loop in the absence of tree-level soft SUSY breaking.  This is
the case in the no-scale model when the PV couplings are chosen so
that the renormalized K\"ahler potential does not break K\"ahler
invariance.  K\"ahler invariance is necessarily broken by gauge
coupling renormalization (unless specific constraints are imposed on
the low energy theory) because there is no similar freedom to adjust
the relevant PV couplings. In the context of string-derived
supergravity, field theory anomalies for K\"ahler transformations
associated with the exact perturbative symmetries of string theory
must be canceled, for example by the introduction of a Green-Schwarz
counterterm~\cite{gs} in the case of gauge coupling renormalization.
This breaks the no-scale structure of the untwisted matter sector, and
there are generally soft SUSY terms at tree level~\cite{bgw}, with
supersymmetry broken in the dilaton ($S$) sector.  (In fact if modular
invariance is not broken by string nonperturbative effects, the moduli
are stabilized at self-dual points with $F^T=0$.)  One-loop effects
can nevertheless be important, especially for gaugino
masses~\cite{bn}.  A general analysis of soft supersymmetry breaking
in these models is in progress~\cite{bgn}.

\vskip .5cm
\noindent {\bf Acknowledgements}
\vskip .5cm We thank Nima Arkani-Hamed, Pierre Bin\'etruy, Andreas
Birkedal-Hansen, Joel Geidt, Hitoshi Murayama and Erich Poppitz for
discussions.  This work was supported in part by the Director, Office
of Science, Office of Basic Energy Services, of the U.S. Department of
Energy under Contract DE-AC03-76SF00098 and in part by the National
Science Foundation under grant PHY-95-14797.

\end{document}